# Challenges of movement quality using motion capture in theatre

**Georges Gagneré**
University Paris 8
2 avenue de La Liberté 93326 Saint-Denis
georges.gagnere@univ-paris8.fr

**Cédric Plessiet**
University Paris 8
2 avenue de La Liberté 93326 Saint-Denis
cedric.plessiet@univ-paris8.fr

**Andy Lavender**
University of Warwick
Millburn Hill Road Coventry CV4 7HS
a.lavender@warwick.ac.uk

**Tim White**
University of Warwick
Millburn Hill Road Coventry CV4 7HS
t.white@warwick.ac.uk

## ABSTRACT

We describe[1] two case studies of AvatarStaging theatrical mixed reality framework combining avatars and performers acting in an artistic context. We outline a qualitative approach toward the condition for stage presence for the avatars. We describe the motion control solutions we experimented with from the perspective of building a protocol of avatar direction in a mixed reality appropriate to live performance.

## 1 INTRODUCTION

In this paper we focus on the generation and performance of avatars, who exist within a conventional screen space (featuring 3D in-world scenography) that must also be understood in co-relation to a three-dimensional theatre space and co-present performers, in a real-time moment of creation and interrelation. This raises questions concerning the 'avatarization' of the theatrical physical body and the emergence of new acting processes, from the point of view of the performing arts and the physical stage. Acting issues are indeed often explored in an autonomous perspective [2] [3]. But it appears that even if the avatar is under human control, the plausibility of her presence remains an issue [10] and requires new approaches to tune interactions [8].

### 1.1 AvatarStaging mixed reality setup

A physical actor, called the 'mocaptor', wears a motion capture geo-spatial low-cost system (fig. 1). She acts in space C and controls an avatar in digital scenery B. The avatar interacts with a performer through a 3D image projected in the rear of space A, in front of the stage director and audience in space E. Transformations in real time of the position of the avatar are performed both by the digital artist in space D, and the 'manipulator' in space E, who uses a gamepad to modify the avatar reference transform T (the avatar's position) in space B.

The manipulator moves the avatar in a forward/backward direction, laterally to the right or the left, and up and down. She rotates the yaw for controlling the movement direction, and the pitch to put the avatar on a horizontal position. She focuses on two main functions:

- adjusting the scenic address of the avatar towards the performer in the A space, from the point of view of the audience; and respecting perspectival constraints [6]
- accompanying the mocaptor in augmented movements that are usually not accessible to a performer, as, for instance, floating in the air.

We have called this hybrid setup AvatarStaging and achieve it using the AKeNe library implemented in a videogame engine for controlling the avatars [7].

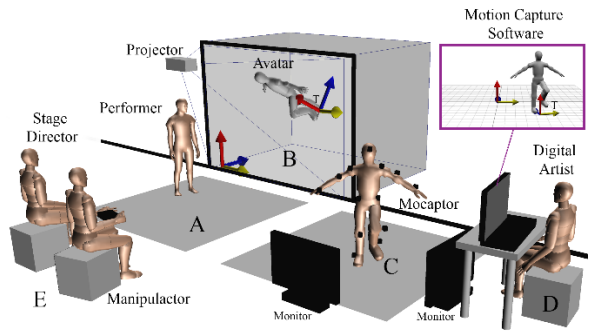

**Figure 1: AvatarStaging mixed reality setup**

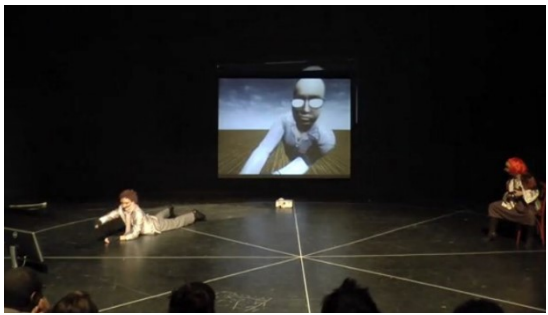

Figure 2: Masterclass in Conservatoire National Supérieur d'Art Dramatique (Paris)

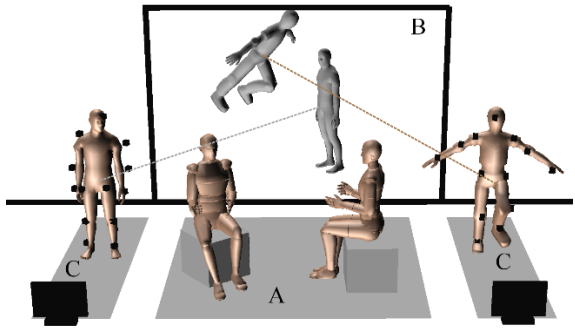

**Figure 3. Configuration of case study N°1**

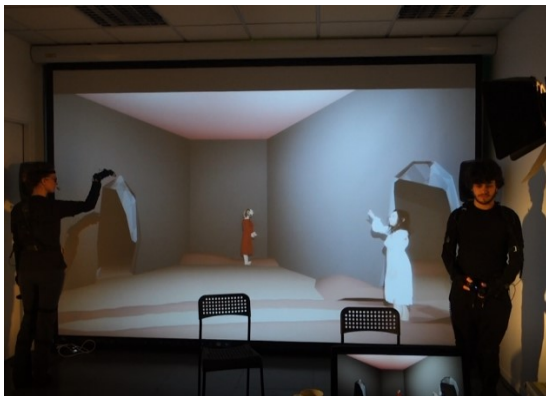

**Figure 4: Case study N°1 - Rehearsals**

## 1.2 Pedagogical and professional use cases

We introduce below the relationships between mocaptors, manipulators, avatars and performers in the context of two case studies. We consider the outcomes of practice-based research conducted by way of a series of interdisciplinary workshops in 2017 as part of a Paris 8 university research project entitled The Augmented Stage, focusing on the development and presentation of theatre and performance in and through digital technologies [12]. The project involved collaboration between software programmers, digital designers and theatre makers and explored the experimental setup with students (fig. 2) and professional artists [13]. Acting issues raised by inhabiting the second body of an avatar are central [1] [9].

## 2 CASE STUDY N°1: DIALOGUE BETWEEN AVATARS AND PEFORMERS

The first case study is based on configuration 1 of the AvatarStaging set up, that shows two corridors C on each lateral side of the stage A with a feedback monitor on the downstage extremities (fig. 3), allowing the mocaptors to see the avatars created in-screen behind them. The mocaptors use the big video projection screen as an alternative feedback when they are walking upstage. This set-up optimizes the visual feedback arrangement.

### 2.1 From a small corridor to a large 3D scenery

The 3D space is larger than the length of the corridors. When an avatar crosses it from one door to the other, she follows this routine: at the extremity of the corridor, she stops her walk, turns, and during this rotation the manipulactor changes the yaw parameter of his avatar in the opposite direction, in order to counterbalance the rotation. The avatar seems to wait in situ. As soon as the U-turn is completed, the mocaptor goes on to her goal (the end-point of this movement trajectory) (fig. 4).

To achieve a smooth and imperceptible counterbalancing, the mocaptor rotation must be precise at the outset, so that the manipulactor does not have to hesitate. It should be slow, allowing time for the manipulactor carefully to control the gamepad thumb stick. The relationship between the two partners was founded on an empathic collaboration geared towards a symbiosis of the avatar movement.

### 2.2 An empathetic collaboration for augmented acting

We know that the mocaptor prefers an empathetic relationship with the image of her avatar on the feedback monitors [5]. That means that she looks at the image and 'translates' the point of view of the avatar: if the avatar faces her on the screen, she will go to her right to have the avatar moving to the left in the image, but to her own right in the relative 3D space. Performing a mental rotation symmetry, the mocaptor is adopting a disembodied self-location to take the avatar visuo-spatial perspective [11]. The manipulactor should develop the same relationship with the avatar through the screen to understand where the avatar must go. When the link is broken for any reasons, it is hard to change the control rules and return to the sympathetic and reflective/symmetrical point of view.

## 3 PLAUSIBILITY AND QUALITY OF MOVEMENTS

### 3.1 Constraints in combining interactions

In this use case, the mocaptor faces three other acting partners: the other avatar inside the 3D scenery and the two physical performers on stage A. An important consideration in conventional theatre is the circulation of attention: who is at the centre of the respective actor's attention, and to whom should other partner actors pay attention? This is decided in common agreement between actors and director, whilst the means of presenting this focus remains a powerful tool of expressivity for the actor.

The manipulactor organizes the system of attention so that the corridor axis provides the address to the performers in space A [7]. The mocaptor ascertains the direction of her avatar partner in space B with the help of the feedback monitor. Indeed, with the audience watching the same image, the mocaptor consequently only needs to imagine around her the digital scenery, respecting the feedback image, and find the position of her partner situated in the same scenery (fig. 5 and 6).

The mocaptor has three choices at any moment: looking in the direction of the corridor axis to have a contact with the onstage performer; looking in the direction of the eyeline that matches the imagined avatar partner; or looking at another point to express a personal feeling deliberately disconnected from her partners, which offers a rich expressive variety. This requires a deep concentration in the continuous collaboration between mocaptor and manipulactor to maintain the integrity and consistency of the movement intentionality of the avatar.

### 3.2 Who is finally controlling the avatar?

Walking within a small corridor necessarily gives responsibility for the movement orientation to the manipulactor – that's to say, an operator who is offstage. We made the same conclusion about the necessity of keeping the scenic address of the avatar in relation to the performers: only the manipulactor can achieve this because of her distance from the stage. However, guiding the mocaptor in this way (at least, by intervening in the movement of the avatar) does not deprive her of initiative. On the contrary, close attention to the avatar's body language helps the manipulactor to understand the intentionality of the movement. Sharpening her avatar body analysis, the manipulactor succeeds in catching the mocaptor's desires that increase the plausibility of the avatar's presence [10].

From the mocaptor's point of view, the difficulty was to maintain a degree of liberty from the defining linear axis of the corridor: when was it pertinent to look aside and break the scenic address built by the manipulactor? Contradictory moves can easily break the spatial continuity of the avatar's acting in relation to her partners and the audience. This remains a live issue that must continually be kept in mind to present a seamless, spatially coherent performance. We are facing a symbiotic but unpredictable improvisation, balanced by a common understanding of the objectives of the scenic actions, driven by the stage director.

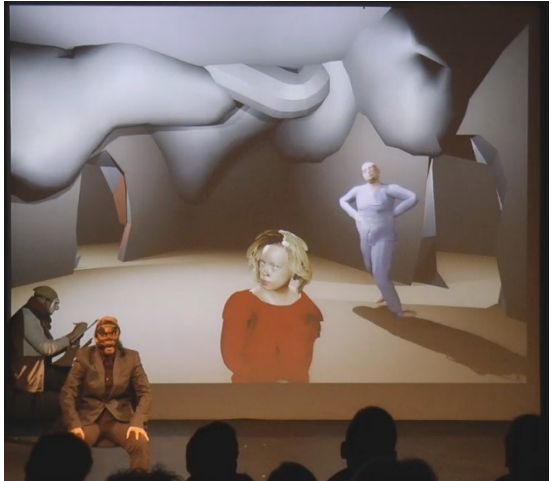

**Figure 5. Case study N°1 – scenic eyelines**

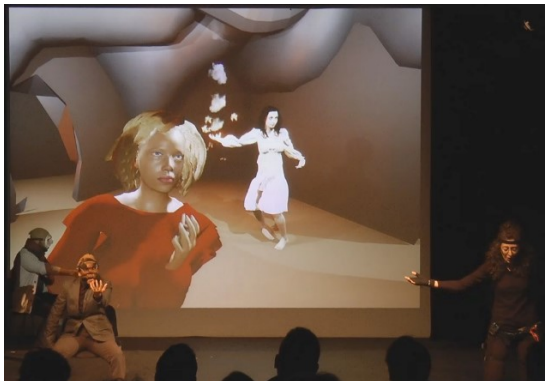

**Figure 6. Case study N°1 - scenic eyelines**

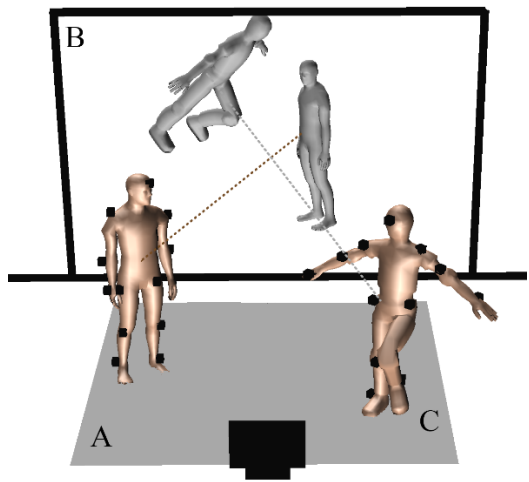

**Figure 7. Configuration of case study N°2**

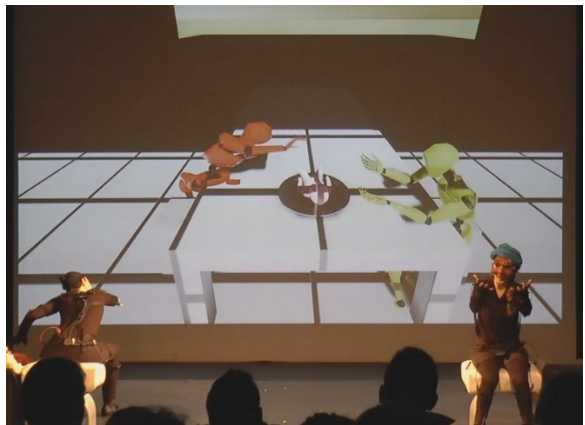

**Figure 8. Case study N°2 – the banquet**

The dialogue between both human controllers seems to be deeper when it is formulated in term of movement quality instead of specific goals [4]. However, the grammar of these new interactions needs to be consolidated.

## 4  CASE STUDY N°2: FUSION OF PERFORMER AND MOCAPTOR

Configuration 2 illustrates the case when spaces A and C are juxtaposed, transforming performers into mocaptors (fig. 7). This is the AvatarStaging configuration set-up for case study N°2 [13], with no side corridors dedicated to the control of the avatar. The performers have one feedback monitor downstage and use the video projection screen upstage for visual feedback when facing away from the audience. This required manipulactors and mocaptors to sharpen their skills to maintain a plausible effect of liveness in the avatars in-screen.

### 4.1  Movement contextualization

Case study N°2 entailed a parallel presentation at any moment. The performers presented their characterization directly to the audience, rather than to each other – that's to say, they did not interact in terms of the normal spatial relationships you would expect between performers onstage; meanwhile they created the movements of the avatars in the virtual scenography, and a plausible (realist-oriented) spatial relationship between the avatars in-screen. This was made possible by experienced mocaptors being able to play two levels of scenic reality at once, and manipulactors able to adapt, on the fly, the avatar movements to the 3D situation.

We presented two dialogue situations in very different contexts: a banquet with two robot figures in a cold white room (fig. 8), and few minutes later a bath with a naked warrior caressed by a woman (fig. 9), played by the same mocaptors. In each situation, we notice how the body of the avatar deeply impacts the way the audience receives the performed actions.

All these different nuances of movement, coming from the actions of two performer/mocaptors who were required to operate within a restricted physical space, offer a rich field of body-movement research concerning relations between live action, characterization, real-time avatar generation and dramatic/narrative development.

### 4.2  Opening dramaturgical possibilities

An interesting consequence of this set-up was to open for the spectator the larger meaning of interrelated body movements, which are distinct but co-created across different kinds of space.  This offers new dramaturgical possibilities both for the stage director and the performer/mocaptor. It led us to the idea of changing the point of view within any particular scene, using the powerful real-time programming features currently offered by the video game engine. Fig. 10 shows four different shots we built to run one scene of the show. This second way of contextualizing avatar movements is also challenging for the mocaptors, who must play virtual eyelines and relationships from an isolated position.

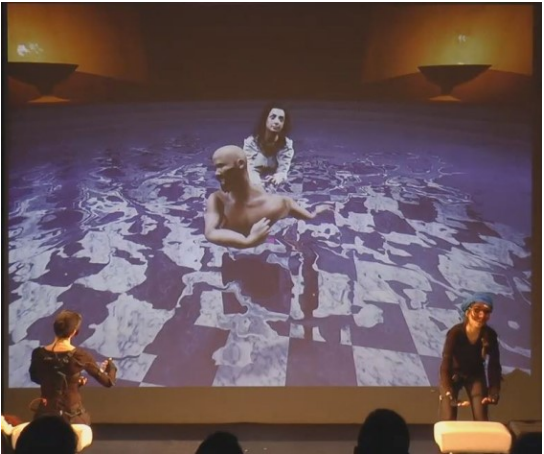

**Figure 9. Case study N°2 – the bath**

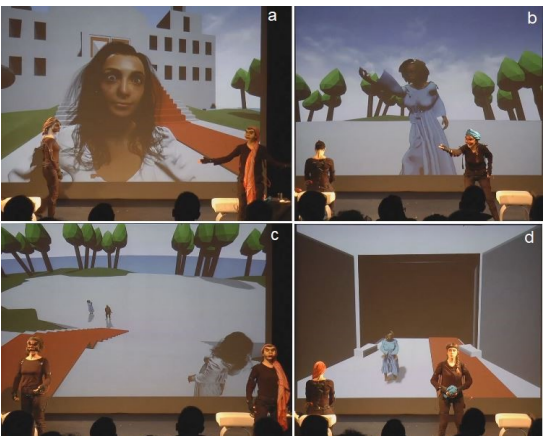

**Figure 10. Case study N°2 – four shots**

In future, we aim to formalize the protocols between the stage director, the mocaptor, the manipulactor and the performer that are needed to establish good practices for avatar direction.

**ACKNOWLEDGMENTS**
This research was made within the framework of "Scène Augmentée" research project, supported by Labex Arts H2H, Investissements d'Avenir French Program, with its partners: Paris 8 University laboratories INRéV and Scènes du monde, création, savoirs critiques, the Conservatoire National Supérieur d'Art Dramatique de Paris, compagnie Gente Gente!, Le Cube - Centre de création numérique, Issy-Les-Moulineaux, Warwick University, and didascalie.net.